\documentclass[aps,prb,onecolumn,superscriptaddress]{revtex4-1}
\usepackage{amsmath}
\usepackage{graphics}
\usepackage{graphicx}
\usepackage{epsfig}
\usepackage{amssymb}
\usepackage{amsfonts}
\usepackage{color}
\usepackage{rotating}
 
\binoppenalty=\maxdimen
\relpenalty=\maxdimen

\begin{document}

\title{Magnetic field sorting of superconducting graphite particles with T$_c$$>$400K}

\author{Manuel  \surname{N\'u\~nez-Regueiro}}
\email{Corresponding author. E-mail: manolo.nunez-regueiro@neel.cnrs.fr }
\affiliation{Univ. Grenoble Alpes, CNRS, Grenoble INP, Institut N\'eel, 38000 Grenoble, France}

\author{Thibaut \surname{Devillers}}
\affiliation{Univ. Grenoble Alpes, CNRS, Grenoble INP, Institut N\'eel, 38000 Grenoble, France}

\author{Eric \surname{Beaugnon}}
\affiliation{Univ. Grenoble Alpes, INSA Toulouse, University Toulouse Paul Sabatier, EMFL, CNRS, LNCMI, 38000 Grenoble, France}

\author{Armand \surname{de Marles}}
\affiliation{Univ. Grenoble Alpes, CNRS, Grenoble INP, Institut N\'eel, 38000 Grenoble, France}

\author{Thierry \surname{Crozes}}\affiliation{Univ. Grenoble Alpes, CNRS, Grenoble INP, Institut N\'eel, 38000 Grenoble, France}

\author{S\'ebastien \surname{Pairis}}\affiliation{Univ. Grenoble Alpes, CNRS, Grenoble INP, Institut N\'eel, 38000 Grenoble, France}

\author{Christopher \surname{Swale}}
\affiliation{Institute for Advanced Biosciences (IAB), Team Host-Pathogen Interactions and Immunity to Infection, INSERM U1209, CNRS UMR5309, University Grenoble Alpes, Grenoble, France}

\author{Holger \surname{Klein}}\affiliation{Univ. Grenoble Alpes, CNRS, Grenoble INP, Institut N\'eel, 38000 Grenoble, France}

\author{Olivier \surname{Leynaud}}\affiliation{Univ. Grenoble Alpes, CNRS, Grenoble INP, Institut N\'eel, 38000 Grenoble, France}

\author{Abdelali \surname{Hadj-Azzem}}\affiliation{Univ. Grenoble Alpes, CNRS, Grenoble INP, Institut N\'eel, 38000 Grenoble, France}

\author{Fr\'ed\'eric \surname{Gay}}\affiliation{Univ. Grenoble Alpes, CNRS, Grenoble INP, Institut N\'eel, 38000 Grenoble, France}

\author{Didier \surname{Dufeu}}\affiliation{Univ. Grenoble Alpes, CNRS, Grenoble INP, Institut N\'eel, 38000 Grenoble, France}

\date{\today }

\begin{abstract}

It has been claimed that graphite hosts superconductivity at room temperature, although all efforts to isolate it have been vain.  Here we report a separation method that uses magnetic field gradients to sort the superconducting from  normal grains out of industrial graphite powders. We have obtained a concentrate of above room temperature superconducting particles. Electrical resistance measurements on agglomerates of sorted grains of three types of graphite show transition temperatures up to T$_{c{_{onset}}} \sim$ 700K with  zero resistance up to $\sim$ 500K. Magnetization measurements confirm these values through jumps  at \textit{T$_c$} in the zero field cooled curves, and by the occurrence diamagnetic hysteretic cycles  shrinking with temperature. Our results open the door towards the study of above room temperature superconducting ill-stacked graphite phases.  
\end{abstract}
\maketitle
\onecolumngrid

\section{Introduction}

After the discovery of superconductivity\cite{Onnes} in mercury by Kamerlingh Onnes in 1911 the quest for superconducting materials has driven research in solid state physics and materials science in the XXth century. Following mercury with $T_c$=4.1~K, metallic alloys of the A15 type structure\cite{NbGe3} have been produced with $T_c$'s of the order of 22~K. But it was not until the late 80s with the discovery\cite{Bednorz} of 2D copper oxides that temperatures as high as liquid nitrogen could be reached. This family supports the highest temperature, $T_c$=166~K, in fluorinated Hg-1223, under a pressure of 26~GPa\cite{Hg1223F}. Lately, hydrogen derived materials in the extremely high pressures of 200~GPa, scratched the room temperature goal\cite{Eremets}. The pursuit of superconducting materials of ever increasing transition temperatures is a very active field of research due to its indisputable benefits in many applications.

For more than twenty years there have been recurrent claims for the existence of unidentified microscopic amounts of ferromagnetism and room temperature superconductivity in different types of graphite, i.e. highly oriented pyrolytic graphite (HOPG) and natural single crystals of AB (Bernal)  and ABC (rhombohedral)  \cite{EsquinaziGraf,EsquinWater,EsquinPersistent}. Its origin has been attributed to granular superconductivity in graphite interfaces\cite{Heikila}, with transition temperature presumedly to be above room temperature. However, experiments aiming at revealing these claims have been very disappointing mostly due to a lack of reproducibility and/or finding very low transition temperatures that makes these materials irrelevant in the context of high temperature performances. One recent example is the Chern ferromagnetic insulating states\cite{Chen}. The associated superconductivity \cite{Cao} has been observed in nano devices made of a few layers of AA graphene twisted at a \textit{magic angle}. The generated Moir\'e patterns can be viewed as adatoms that generate extremely flat bands\cite{Mayou, MacDonald}. Although the superconducting transition temperatures for these systems are in the range of liquid helium, these results stress the need of a deeper inspection of the possible superconducting properties of all types of graphene stacking materials. To make things worst, if \textit{magic angles} can explain the occurrence of superconductivity in bulk graphite the number of grains having the appropriate geometry would be really tiny.

Recently, superconducting transitions with several high temperature $T_c$'s, 110~K, 245~K, 320~K, have been reported in KC$_8$ deintercalated at room temperature\cite{Layek}. The low kinetics of such a transformation induces a bulk twisted graphite geometry leading to different twist angles and dopings. As mentioned above, only a small fraction of crystallites have the superconducting properties. Superconductivity in these compounds is thus granular, although almost percolation has been found for the transition at 245~K. During these studies, we observed that the virgin graphite (HOPG or Grafoil) used as starting product for the intercalations also showed superconducting-like diamagnetic cycles, that shrank to zero with increasing temperature towards a $T_c\sim$ 550~K\cite{Raphael}.

Having these results in mind, we have successfully applied a magnetic separation methodology to three different commercial, off-the-shelf, forms of graphite: synthetic (SynG), natural (NatG) or spectroscopically pure (PurG)  (see Methods). Graphite is diamagnetic with a magnetic susceptibility that is roughly of the order of $-10^{-5}$ to $-10^{-3}$ whereas an ideal superconductor has a susceptibility of $-1$. The driving force of this separation is the interaction between a diamagnetic material and an inhomogeneous magnetic field which tends to push the diamagnetic material away from the source of magnetic field. This magnetic force being proportional to the magnetic field, the gradient of magnetic field and  the magnetic susceptibility of the material, the graphite powders that are superconducting at room temperature can be separated from  regular graphite in the presence of a magnetic field of large amplitude and high spatial gradient. Thanks to this purification method, features related to superconductivity are being enhanced and the error bars in the measurements are smaller, leading us to conclude beyond any doubt that high temperature superconductivity in \textit{some form of} graphite is a real, reproducible property. All these analysis point towards a high temperature $T_c$ at around 550-600~K in magnetically sorted synthetic, natural or spectroscopically pure graphite powders.  As the amount of material is tiny the graphite allotrope responsible for the superconductivity has not been identified, yet.

We describe in this manuscript the separation technique and the characterization of basic superconducting properties, i.e. electrical resistance and magnetic properties of the concentrate of superconducting particles.

\section{Results}

\begin{figure}
	\centering
	\includegraphics[width=16cm]{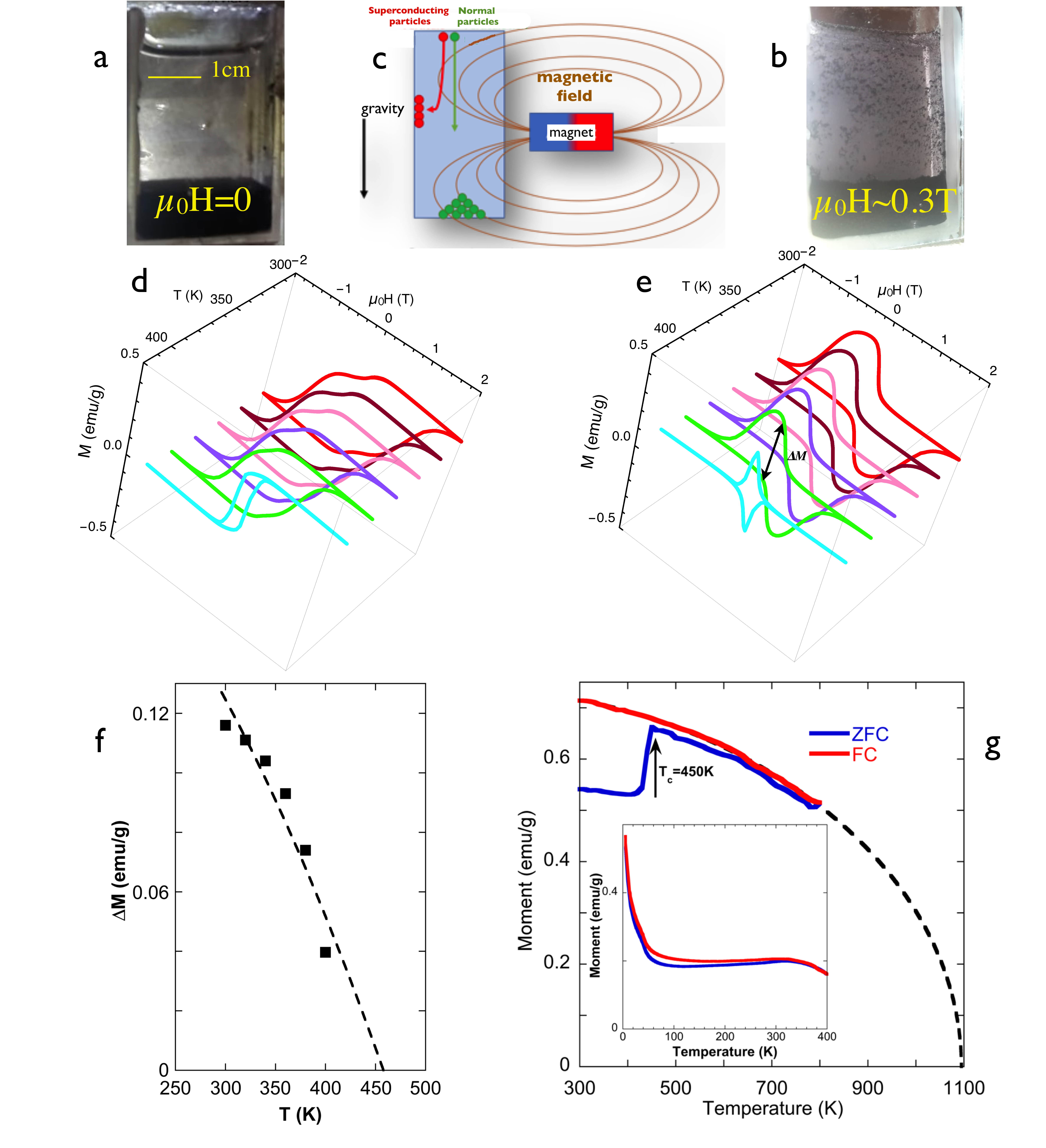}
	\caption{(a)  suspension of graphite particles  with MnCl$_2$  after decantating in absence of external magnetic field: the flask faces are clear. (b)same solution after decantating in the vicinity of a bulk magnet. Carbon particle flakes are stuck on the window by the magnetic field gradient.(c) Schematics of the process. (d) Magnetization cycles for a MVS SynG sample measured in the MPMS3 for different temperatures above 300~K. (e) Same cycles after substraction of a ferromagnetic sigmoidal cycle fitted on the 400~K cycle. (f) Variation with temperature of the vertical spread of these last cycles at 0.3T. A fit with expression (\ref{Jc}) yields a \textit{$T_c$}=455$\pm$40~K. (g)  ZFC and FC measurements above 300~K for a MVS SynG sample in the MPMS3 oven-stick at 1T. We observe a superconducting jump at 450~K compatible with the \textit{$T_c$} obtained from figure f. Insert: ZFC and FC measurements for a same batch sample below 400~K. We observe a paramagnetic increase of magnetization at low temperatures.}
	\label{System}
\end{figure}

As the obtention of superconducting particles is central to this work, we describe here the Magnetic Decantation Separation (MDS) method first used for sorting superconducting from regular graphite grains. It relies on the differential migration of  particles of different magnetic susceptibilities upon decanting under a magnetic field gradient. A suspension of graphite particles in a  paramagnetic solution containing MnCl$_2$ and tannic acid  is left decanting in a spectrophotometric cell for typically 12 hours (Fig.~\ref{System}c). Alone, all particles end up at the bottom of the cell (Fig.~\ref{System}a). Against a magnet, the particles with highest susceptibilities are accumulated on the face opposed to the magnet (Fig.~\ref{System}b). These particles are then extracted and cleaned.

We now discuss our magnetization data. On Fig.~\ref{System}d we show  magnetization cycles at different temperatures up to 400~K of a large (3mg) MVS SynG sample.
We observe that the cycles diminish slightly with increasing temperature and that the last cycle has shrunk considerably. The profile has changed from a complex and wide shape to an almost anhysteretic ferromagnetic form, suggesting that there is a transition around  $\sim$ 400~K. We therefore  fitted a sigmoidal ferromagnetic cycle on the data at 400~K and subtracted it from the lower temperature cycles.

The cycles shown on Fig.~\ref{System}e are typical of a superconducting diamagnetic hysteresis\cite{YBCO-Mariposa} . As done previously \cite{Layek, Raphael}, we plot on Fig.~\ref{System}f the $\Delta$M(0.3T) of each cycle as a function of temperature  and fit it with expression (\ref{Jc}).
\begin{equation} \label{Jc}
{J_c(T)=J_{c0}[1-(T/T_c)^ 2]}
\end{equation} The result yields a $T_c\sim 455\pm$40~K (Fig.~\ref{System}g). 

Fig.~\ref{System}g shows the ZFC and FC magnetization temperature dependences at 1T for a MVS SynG, after annealing at 800~K. The overall shape can be fitted by the phenomenological expression (\ref{Ferro}) used to describe the evolution of ferromagnetic transitions \cite{Carley}(black dashed line on figure). 
\begin{equation}
{M_s(T,H)=M_0(H)[1-(T/T_0)^2]^{1/2}  }
\label{Ferro}
\end{equation} 
We obtain a \textit{T}$_{\mathrm{Curie}}$=1095$\pm$50~K value that suggests the presence of metallic iron impurities. On the ZFC curve we observe a clear diamagnetic transition, that is not detected on the FC curve, as expected for a superconducting transition. The observed \textit{T}$_c\sim$450~K is in agreement with the one obtained from the shrinking cycles. Also these values of $T_c$ are recurrent in the electrical resistance measurements of MDS or MVS SynG (see e.g. SI Fig.~\ref{Res-SI}a).

  The zero field cooled (ZFC) and field cooled (FC) magnetization at 1T between 400~K and 5~K are shown the insert of Fig.~\ref{System}g. A rather flat dependence with (\textit{T}$_{\mathrm{Curie}} \sim$ -4~K) Curie-like variation below $\sim$ 50~K.

We present now some examples of the electrical resistivity measurements performed in  agglomerates of graphite particles obtained from different types and methods of concentration. More are presented in SI Fig.~\ref{Res-SI}.

\begin{figure}
	\centering
	\includegraphics[width=16cm]{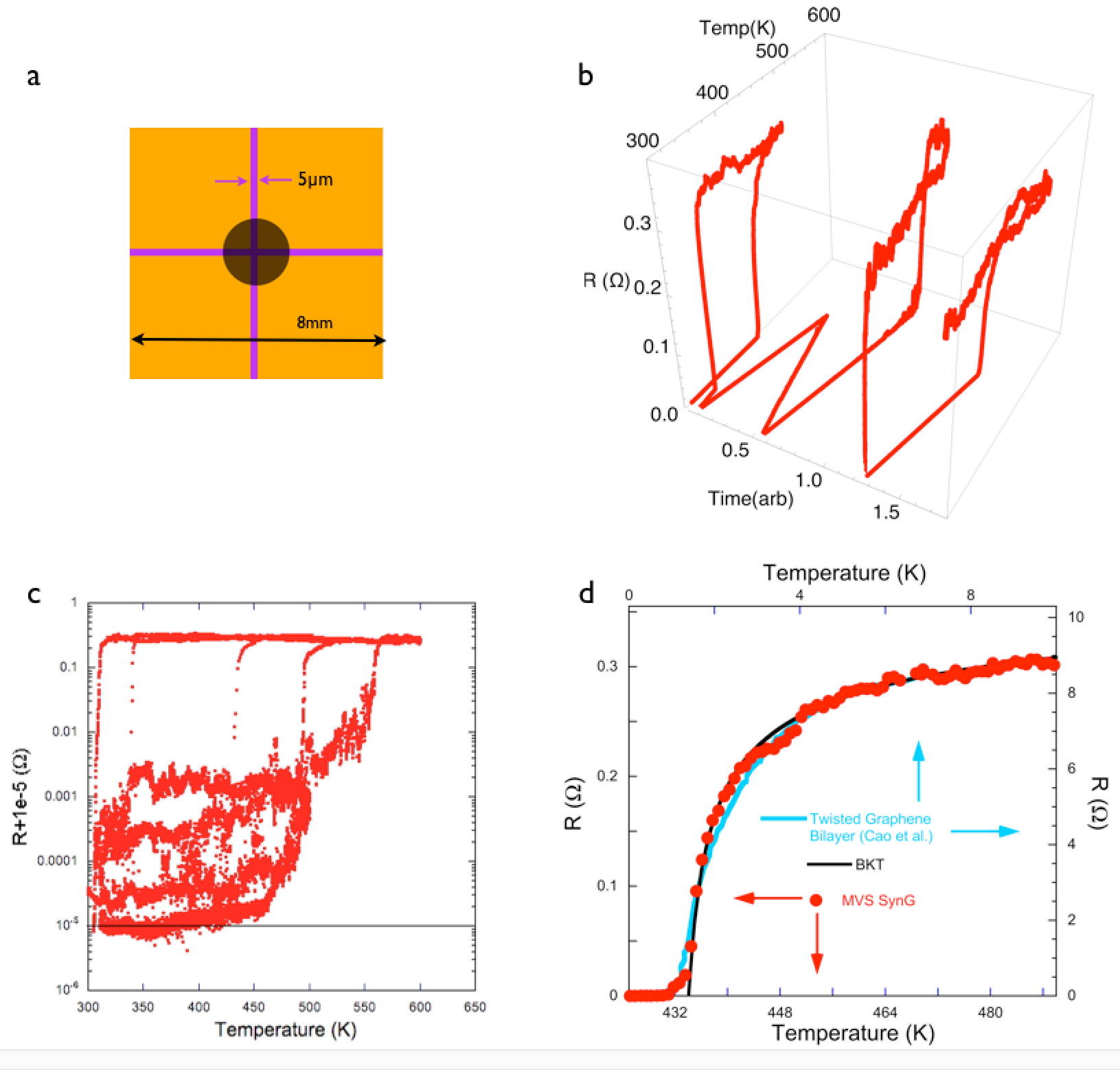}
	\caption{(a) Device for measuring the resistance, dimensions 1x1cm, used for van der Pauw method measurements, the width of the gap is 5$\mu$m. (b) 3D plot of the electrical resistance of a small granular MVS SynG sample, as the temperature is cycled. We observe that in the second cycle up to 500~K the sample remains always superconducting.  (c) Logarithmic plot of the resistance to evaluate the degree of zero-resistance where a residual value of 1e-5 $\Omega$ has been added to the original data. The zero is marked by the black line. Insert:  A drop (small black circle) of a sorted graphite suspension was allowed to dry in the middle before measuring. (d) One of the transitions in (b) (red dots)has been fitted (black line) by the Halperin-Nelson formula for a BKT superconducting transition. We also compare the curve to the transition\cite{Cao} of twisted bilayer graphene at low temperatures (cyan line) on a different temperature scale.}
	\label{Res-BKT_PNAS}
\end{figure}

The  electrical resistance on a MVS SynG sample, measured in a van der Pauw method device, is shown in Fig.~\ref{Res-BKT_PNAS}a. The sample was cycled between 300 and 500 (or 600)K. The first cycle that showed a high resistance activated behavior is not shown for clarity. The transition temperatures obtained upon increasing temperature are higher than those while decreasing temperature. There is a cycle where the sample stays in the superconducting state up to 500~K. Albeit the normal state is very noisy.

All of the transitions observed in this sample have a rounded shape that falls sharply at $T_c$, very similar to those reported at $\sim$1~K in twisted graphene bilayers\cite{Cao}, attributed to a Berezinskii-Kosterlitz-Thouless (BKT) transition.  
Thus we show, on Fig.~\ref{Res-BKT_PNAS}d one of our transitions fitted with the Halperin-Nelson\cite{HalperinNelson} expression, R $\sim$ exp(\textit{b}/(\textit{T}-$T_c$)$^{1/2})$, corresponding to a superconducting transition of 2D BKT type, where \textit{b} is a parameter related to the vortex core energy. In our measurements, this type of BKT behavior is found using the smallest gold contact devices. It is unstable and difficult to observe.  We also compare it to one superconducting transition in a twisted graphene bilayer device\cite{Cao} (cyan line  Fig.~\ref{Res-BKT_PNAS}d).

On Fig.~\ref{Aslamazov}a the standard Aslamazov-Larkin\cite{Aslamazov} expression, $\Delta \mathrm{S}=B[(T-T_c)/T_c]^{-\delta}$ is used to extract the dimensionality of the fluctuations for three of these transitions. $\Delta \mathrm{S}$ is the increase of conductance above $T_c$ and $B$ is a constant. It yields $\delta\cong$ 1, implying strictly two dimensional superconducting fluctuations down to $T_c$, within the precision of our measurement.

On Fig.~\ref{Res-3D}b we show the resistance of a MHS NatG sample mounted on the device shown in  Fig.~\ref{Res-3D}a . The transition shapes are very different from those of the preceding sample. They are more gradual, more extended and $\textit{S}$ -shape. The value of the zero resistance is zoomed on Fig.~\ref{Res-3D}c. We show on Fig.~\ref{Res-3D}d one of the cycles. The onset $T_c$ occurs at 700~K for increasing temperature, with a middle value of 650~K. In any case, these values are higher than those of the BKT transitions.

Finally we analyze the fluctuations by the Aslamazov-Larkin method (Fig.~\ref{Aslamazov}b). At higher temperatures a linear $\delta\cong$ 1, 2D behavior is observed as in Fig.~\ref{Aslamazov}a. However, when the sample approaches $T_c$,  the slope decreases to a $\delta\cong$ 0.5, i.e. the fluctuations become 3D, before attaining a critical regime with $\delta\le$ 0.1 near to $T_c$. We thus obtain a 3D transition instead of the pure 2D type of Fig.~\ref{Aslamazov}a, explaining the difference between the two types of resistance curves.

Results show clearly that superconductivity can be stable up to a maximum of $\sim$ 700~K in our samples. On the other hand, the hysteresis is typical of a weak link network\cite{Felner}, although other possibilities remain open.

\section{Discussion}

Our separation method has allowed to increase the amount of superconducting grains in graphite materials although there is still a variable amount of non-superconducting graphite in our samples. One explanation is that in our solutions the superconducting particles attach to a certain amount of non-superconducting particles. Another possibility is that each sorted grain is only partially superconducting, as expected if the origin of this high temperature superconductivity is due to ABC-AB  interfaces or twisted layers. It is therefore difficult to estimate the percentage of superconducting particles. For the samples displaying zero resistance transitions, we observe percolation which according to the Landauer percolation criterion\cite{Landauer}  implies $\sim$ 30\% of superconducting grains.

On the other hand, all of our samples display a ferromagnetic cycle convoluted with the diamagnetic one. A similar situation exists in UCoGe\cite{Carley}, with a ferromagnetic state appearing well above superconductivity.
While our ferromagnetic phases seemingly appear at the same temperature as superconductivity (Fig.~\ref{TDA1}f of SI). Presumably, they may be intrinsic, as has been shown in HOPG\cite{Hebard}. Ferromagnetism has been measured in twisted bilayer graphene\cite{Chen}, but at a region of doping different to the one corresponding to the superconducting phase.

Also, all the magnetization versus temperature measurements  present the passage to a paramagnetic phase at low temperatures, even in  spectroscopically sorted pure AGAR powder (Fig.~\ref{TDA1}a and b of SI). As this is not the behavior of a Bernal graphite, it supports the idea that we separate a new type of graphite, probably defective.

A transition temperature as high as 700~K for a variety of graphite can be surprising, considering that the highest $T_c$ obtained with twisted graphene stackings is around 10~K. This paradox can be resolved if we consider that superconductivity originates through a mechanism based on the presence of \textit{flat bands}, as expected for ABC-AB or twisted multilayers interfaces.  The dispersionless energy spectrum of flat bands has a singular density of states, so that $T_c \sim \lambda$, instead of being $T_c \sim \exp(-1/\lambda)$, where $\lambda$ is the coupling constant. In other words, $T_c$ is practically unbound\cite{Volovik}.  Recent works on flat band systems have shown that their conductive and superconducting properties are not only stable against disorder, but that they can even be strongly enhanced by it\cite{Bouzerar,Thumin}.

The pure superconducting portions present a 2D BKT superconductivity. Moir\'e superconducting bilayer graphite also shows BKT superconducting transitions (see Fig~\ref{Res-BKT_PNAS}d), and they have been studied in detail\cite{Cao}. Theoretical calculations\cite{Julku} for the Moir\'e have shown that the gap mean field $\Delta$ appears at a mean field transition temperature $T_{c}^{\mathrm{MF}}$. However,  2D vortex fluctuations hinder the development of zero resistance state. At a temperature $T_{c}^{\mathrm{BKT}} < T_{c}^{\mathrm{MF}}$ vortex-antivortex pairs are formed and  pinned down, allowing the resistive transition to take place. According to calculations\cite{Julku},  $\Delta$  can be between 5 to 10 times the actual measured $T_{c}^{\mathrm{BKT}}$. We believe that these theoretical analysis are pertinent to our results and, using the same ratio, we should expect for our samples a value of $T_{c}^{\mathrm{MF}}$ of several thousand Kelvin. 

Our work has revealed that there are clearly two type of superconducting transitions: the sharp 2D BKT  and the smeared, often higher in temperature, 3D ones that follows from the Aslamazov-Larkin analysis (Fig.~\ref{Aslamazov}a and b). In addition, there is very likely some kind of proximity effect among superconducting and normal portions. It is tempting to think that the superconductivity induced in the normal and 3D regions might have a higher $T_c$, explaining our results.

In spite of previous claims\cite{EsquinaziGraf}, the measurements here presented are the first clear evidence that superconductivity exists in graphite well above room temperature, regardless of the commercial source of the graphite. Its origin, be it ABC-AB defects\cite{Heikila}, multilayer Moiré stacking faults or other, remains to be demonstrated. Transmission electron microscope (TEM) images from the samples used in Fig.~\ref{Res-3D}b are displayed in Fig.~\ref{Aslamazov}c. The image shows several particles with arbitrary orientation and some of the grains exhibit a stripe contrast showing that there are 2-dimensional domains with a thickness of a few nanometers. The grains showing no contrast could still have such a domain structure, but due to a different orientation these domains might not be visible in this image. For lack of another interpretation, we can only hypothesize that SC may be due to ABC-AB or multilayered twisted defects embedded in a Bernal matrix (See Insert Fig.~\ref{Aslamazov}c). Further crystallographic studies are under way.

The observed $T_c$'s$ \sim$500~K are well in agreement with those previously reported\cite{Raphael} (and at the origin of this research), that were extracted by the shrinking of diamagnetic hysteresis cycles with increasing temperature on off-the-shelf HOPG and grafoil. They also are a strong support towards previous reports for superconductivity in bulk twisted graphite phases\cite{Layek} at lower $T_c$'s, 100, 245~K and 320~K. 

Very important, the fact that we are able to measure zero resistance on an agglomerate of superconducting-normal particles is of capital importance for applications. Graphite particle paints such as Aquadag\textsuperscript{TM}, have been used for a long time in industry for, e.g., painting the interior of cathode tubes to close the electronic beam circuit. We have separated Aquadag grains from the liquid paint by centrifugation, and measured the magnetically sorted ones. On Fig.~\ref{Aslamazov}d  we disclose a preliminary resistance measurement on magnetically sorted particles extracted from Aquadag showing a superconducting transition. It suggests that, if the Aquadag paint instead of being fabricated from normal graphite particles is made from the sorted superconducting ones, we would obtain a superconducting paint whose resistance might be possibly zero above room temperature allowing for the design of superconducting circuits at room temperature. The next step towards applications is to obtain large quantities of superconducting grains, e.g., either by modifications of the Acheson process for synthesizing graphite or by upscaling the separation process to industrial level.

\section{Material and Methods}{\textbf{Graphite sources}:
	four different commercial graphite materials were tested upon magnetic separation to extract a superconducting concentrate (SI Fig.~\ref{X-rays}): (i) Sigma-Aldrich  synthetic powder 282863, grain size $<$ 20 microns (SynG), (ii) Prolabo grain size $<$ 5 microns of natural origin (NatG), (iii) graphite rods from Agar, the purest available graphite, spectroscopically pure carbon, that were ball-milled to obtain the powder(PurG) and (iv) Aquadag \textsuperscript{TM} that was first diluted with water and centrifuged to extract the graphite particles.

	\textbf{Magnetic Decantation separation (MDS)}: Graphite powder is mixed with DI water (2~mg/mL), tannic acid and MnCl$_2$ (1~g/mL). Since graphite is hydrophobic, tannic acid is used as a defloculating agent, prohibiting the agglomeration of particles and slowing down considerably the sedimentation. The role of MnCl$_2$ is to turn water into a paramagnetic solution, and therefore increase the susceptibility contrast between graphite and the solution. The solution is introduced in a  spectrophotometric cell of dimensions 50x40x10~mm with a NdFeB magnet of dimensions 40x40x10~mm in contact with the sidewall of the cell with the direction of magnetization horizontal  and pointing to the cell.  The measured magnetic induction at the surface of the permanent magnet is $\sim$~0.5T. After being stirred, the liquid was  black and opaque. In the absence of the magnet, all particles fell to the bottom by gravity and the solution got clear (Fig.~\ref{System}a). However, with the magnet a small amount of them stayed stuck to the wall opposite to the magnet (Fig.~\ref{System}b). These particles that have supposedly the highest susceptibilities are then  extracted and cleaned from MnCl$_2$ compound using a dialysis cassette dilution process if the quantity was small or a high speed centrifuge with larger amounts.
	
	\textbf{Magnetic Vertically Sorting (MVS)}: is based on a microfluidic devices in which two micro-channels are put in contact on top of each other above a series of alternate up/down millimetric  permanent magnets. A suspension of graphite particles in deionized(DI) water with tannic acid is flowing through the bottom channel while  DI water is introduced in the upper channel. The particles with higher diamagnetic susceptibilities will be repelled further and migrate to the upper channel from which they can be extracted for analysis (see SI Fig~\ref{System_SI}a). This method appears to be more selective than the first one in the sense that the purity of the sorted particles is more important. 
	
	\textbf{Magnetic Horizontally Sorting (MHS)}: is also based on two communicating micro-channels, one containing the graphite suspension and the other DI water. The difference with the previous method is the planar geometry (instead of vertical) and the magnetic landscape which promotes the horizontal migration of high susceptibility particles from one channel to the other(see SI Fig~\ref{System_SI}b). Because of its geometry this system allows for the highest purity of the particle concentrate. More details are given in the Supplementary Material.

	We have performed a total of more than 30 purifications, about 10 for each method. 
	In the microfluidic separation method the suspension was prepared the day before and left decanting overnight to allow only the smallest grains to remain in suspension, $<$1$~\mu$m, with a final sorted mass ranging between 1mg and 0.1mg. We have also tried sorting Aquadag particles centrifuged from the paint. Only the MHS method was successful as the Aquadag particles are extremely small, $\le$100~nm. They did not decant and easily changed channels in the MVS method, even without magnets.
	
	\textbf{High temperature resistivity measurements} were performed in home-made fast optical furnaces, in a four lead DC measurement with four tungsten elastic fingers to ensure the electric contact. Two different apparatus were used, one with four fixed  aligned contacts and the other with four fixed contacts in a van der Pauw configuration. We have measured the resistance using a Keithley 6220 precision current source and a Keithley  2182 nanovolt meter. Direct and reverse current are applied alternatively to remove offsets due to amplifiers and thermometric effect. Typically we inverted current ten times, waiting several seconds for current stabilization and performed an average. The elliptical halogen lamp used for heating was controlled through TDK lambda power supply, and temperature measured with a thermocouple.
	
	Since  samples were made out of grains of very small size it was not so easy to place conveniently on the contacts. Their thin platelet shape together with the high conductance anisotropy of graphite renders zero resistance difficult to attain.  Different types of four gold contacts supports were fabricated, the smallest one by electronic lithography had 100nm wide contacts, with a length of  2mm and separated by 150nm.  A water suspension of the sorted particles was deposited in the middle of the channels with a micro pipette. On drying, surface tension pushed the grains towards the contacts. Sometimes  the dried drop was covered with corn starch or acrylic paint, or water-glass, to increase the gold-grain contacts, and to prevent the particles from flying away under vacuum or inert gas flow. Only about 10\% of our samples show zero resistance, some (40\%) no transition at all, and some (30\%) show a negative resistance below $T_c$. Similar negative resistance results were observed , e.g. in twisted superconducting nanowire yarns\cite{Dongseok-NegRes}. Others (30\%) developed an insulating state at the expected $T_c$, as in granular superconductors with insufficient superconducting percentage\cite{Goldman}.  In all cases it was necessary to anneal the sample at about 700~K or more to obtain a clear transition, probably to eliminate impurities such as tannic acid. Typically, we cycled five times at 10~K per minute increasing the current one order of magnitude in each cycle starting from 1nA. Unfortunately, the cycling to such high temperatures often finished by desegregating the sample before reaching a proper measurement.
	
	\textbf{Magnetization measurements} up to 400~K were performed in a Quantum Design MPMS3,VSM-Squid Magnetometer. The sample holders were thin plastic straws provided by Quantum Design. Quantum Design manufactures also a consumable oven stick for measurements between 300~K and 1000~K. However, its  background signal was very large and, moreover, changed with cycling, allowing us to do few reliable measurements in its short life.}

\textbf{Authorship contribution statement}
MNR: Conceptualization, Sample fabrication by separation, Magnetization and electrical resistivity measurements and analysis, Methodology, Writing original draft, Management.
TD: Development of microfluidic devices for separation, sample fabrication.
EB: Development of static separation trials.
AM: First successful obtention of magnetic separation.
TC: Design and fabrication resistivity devices.
SP: MEB characterization
HK: TEM measurements
CS: Support on sample clean-out techniques.
OL: Support X-rays
AH: Chemistry technical support. 
FG: Optical ovens for electrical resistivity
DD: Support magnetization.

\textbf{Acknowledgments}

 We strongly acknowledge the firm steadfast support of our CNRS and University laboratories. MNR thanks J.E. Lorenzo, M-A. M\'easson and F. Levy-Bertrand for manuscript optimization and J-L. Tholence, K. Hasselbach and C. Paulsen for discussions.

\appendix
\section{SUPPLEMENTARY INFORMATION}
\subsection{Separation Methods}
.

\textbf{Vertical magnetic separation (VMS)}: In this method, we use a microfluidic device made of 5 layers of polymer and double side adhesive stuck on top of each other which define 2 microfluidic channels and a chamber allowing the communication between the channels. The thickness of the channels is defined by the thickness of the double side adhesive, here 140~$\mu$m, and the spacer between the channels is of 100~$\mu$m. The channels are cut in the adhesive by laser cutting and the width is 4~mm. The solution of graphite (10~g/L) with tannic acid ($\sim20$~mg/L) is fed into the lower channel while DI water is fed into the upper channel(Fig.~\ref{System_SI}a). The flow rates are in the range of 50~mL/hour and induced by an overpressure of a few mbar. Considering the size of the channels, the flow can be considered as perfectly laminar, and with a slight overpressure in the upper channel, we can ensure that the graphite solution doesn't diffuse through the upper channel. Multiples NdFeB magnets of dimension 2x5x10~mm with alternate magnetization pointing up or down are placed just below the channels, to maximize at the same time the magnetic field and its gradient to produce the strongest repulsive force along the $z$ direction. Though more reliable and selective than MDS, this technique suffers from imperfections in the channel design and formation of micro-bubbles which disturb the laminar flow and favour the migration of random particles in the upper channel. This technique has the advantage of producing forces significant enough to avoid using MnCl$2$ and skip the last purification step of the previous method. However the output solution is so diluted that it looks like pure water and needs to be evaporated or centrifuged to extract the particles.

\textbf{Horizontal magnetic separation (HMS)}: this technique is based on a single microfluidic channel (200~$\mu$m thick, 4~mm large, 4cm~long) with two inlets (DI water and graphite solution) and two outlets (DI water+high $\chi$ particles and graphite solution). A magnetically soft iron-silicon foil (300~$\mu$m thick, 2~cm large, 4~cm long) is embedded in the base of the micro-channel, with its edge just below the channel going diagonally from one inlet to the opposed outlet(Fig.~\ref{System_SI}b). The system is placed on top of a 50x50x25~mm NdFeB magnet magnetized vertically. The high field gradient produced by the edge of the foil is repelling vertically and  transversely the particles of higher susceptibility from the graphite channel to the DI water channel. This method showed the highest selectivity because the larger distance separating the two channels reduce the risk of unwanted migration related to defects in the system.

\subsection{Materials}

Four different commercial graphite materials were tested upon magnetic separation to extract a superconducting concentrate (SI Fig.~\ref{X-rays}): (i) Sigma-Aldrich  synthetic powder 282863, grain size $<$ 20 microns (SynG), (ii) Prolabo grain size $<$ 5 microns of natural origin (NatG), (iii) graphite rods from Agar, the purest available graphite, spectroscopically pure carbon, that were ball-milled to obtain the powder(PurG) and (iv) Aquadag \textsuperscript{TM} that was first diluted with water and centrifuged to extract the graphite particles. 

We show on SI Fig.~\ref{X-rays} MEB images of the four types of graphite used. We approximately verify that the SynG had been sieved through a 20$\mu$m mesh and that its platelets are larger than those of NatG and PurG. Aquadag grains are very small and thin, corresponding probably to expanded graphite.

The X-ray spectra for SynG shows neat peaks. In the zoom in insert, it is clear that there is a small peak corresponding to about 10\% of rhombohedral graphite. The NatG spectra shows a large amount of silicate contamination, due probably to its natural mining origin. While PurG has a clear turbostratic structure, that presumably will yield many stacking faults.

\section{Results}

\subsection{Electrical Resistance}

We present now some of the electrical resistivity measurements performed in  agglomerates of graphite particles obtained from different types and methods of concentration.

 On Fig.~\ref{Res-3D}a we show a transtion for a MVC SynG. It is the first superconducting transition observed in our study. It has a clear transtion at $\sim$ 450K but no zero resistance. Cycling in temperature yielded slightly different T$_c$'s. The measurement was unstable as particles were removed by the vacuum till total destruction. Results on two MVS NatG samples are shown on Fig.~\ref{Res-3D}b and c. The hysteresis is always present and the onset T$_c$'s on increasing temperatures seem to be higher than 600K, while the transitions are larger with lower temperature middle T$_c$'s. Fig.~\ref{Res-3D}d shows the electrical resistance of a MVS NatG sample showing up and downwards transtion hysteresis. At 650K we do not attain the normal state. On Fig.~\ref{Res-3D}e we show the measurement on a MHS spectroscopically pure AGAR graphite. The onset T$_c$ appears now above 800K.  
 
 \subsection{Magnetization measurements}
 
\textbf{Spectroscopically Pure Graphite}: We  describe the measurements performed on a MHS PurG sample. The sample was small (0.6mg) and could not be measured against the addenda of the oven stick.
 
 We show on Fig.~\ref{TDA1}a a ZFC magnetization measurement as a function of temperature. The signal is clearly diamagnetic and decreases its magnitude as the temperature is increased. Like for the sample showed on the insert of Fig.1g of the main section, at low temperatures we have a Curie-like behavior, with \textit{T$_{Curie}$} $\sim$ -4K. In the case of this spectroscopically pure sample, we would be inclined to accept an intrinsic origin. We also show the corresponding measurement for the milled powder before magnetic sorting. We see that the sorted sample has a diamagnetism more than twice as large at room temperature, confirming the effect of the magnetic field separation. 
 
 The sample being too small to be passed in the MPMS oven-stick, we measured the ZFC magnetization dependence for different fields. The shielding increases with the applied field amplitude besides the upturn at low temperature. As seen on Fig.~\ref{TDA1}b, all the curves converge at high temperatures to the same value, 480$\pm$30K, strongly suggesting that it corresponds to a superconducting transition.
 
\begin{figure}
	\centering
	\includegraphics[width=16cm]{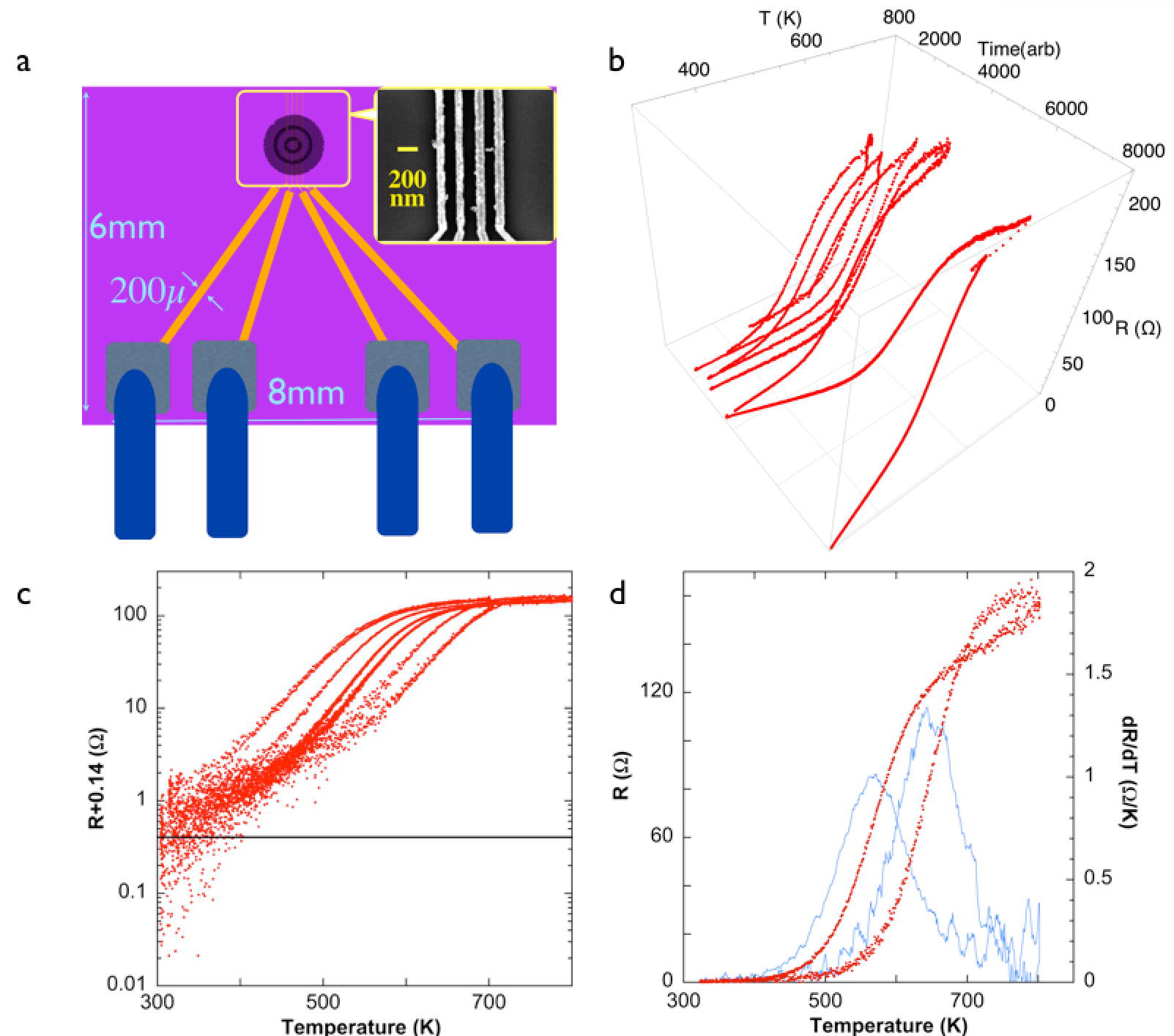}
	\caption{(a) Device for measuring the resistance fabricated by electronic lithography, dimensions 1x1cm. The width and separation of the four small gold strips is 100 and 150nm, respectively. We also schematised the tungsten wire fingers (blue) placed on the dried silver paste contacts (gray) used to make the contacts. (b) 3D plot of the electrical resistance of a sample obtained from MHS NatG, as the temperature is cycled. The onset T$_c$ seems bigger, but the zero resistance appears at lower temperatures. No BKT behavior is observed. (c) Logarithmic plot of the resistance to appreciate the degree of zero-resistance. We have added 0.14 $\Omega$ to put it on the semilog plot. The zero is then marked by the black line.(d) The penultimate cycle is shown in detail, and from the temperature derivative of the resistance, the middle of the transition is determined to be at about 650K on increasing the temperature.}
	\label{Res-3D}
\end{figure}

\begin{figure}
	\centering
	\includegraphics[width=15cm]{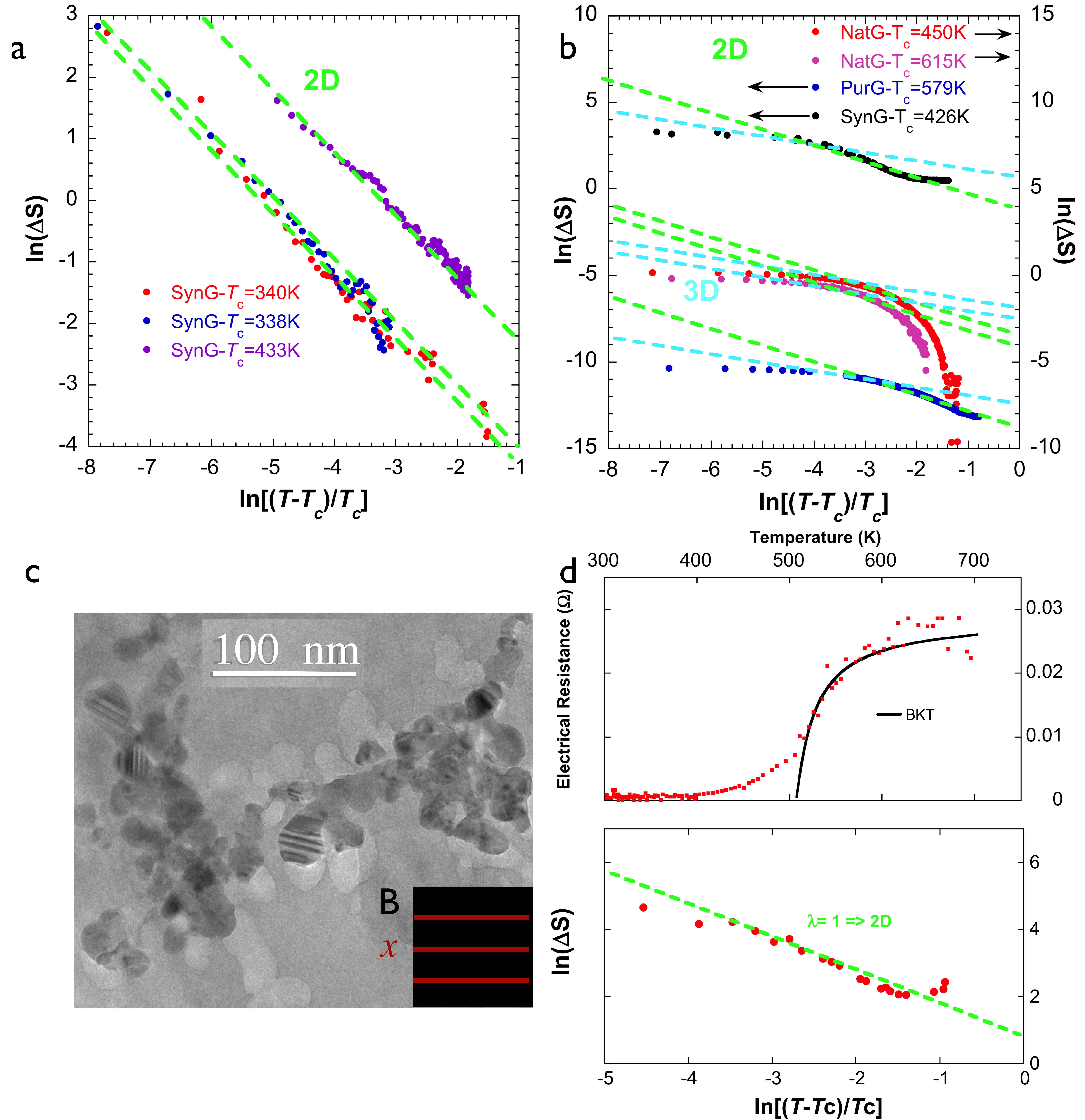}
	\caption{(a)  Aslamazov-Larkin plot of the BKT transitions. The slope $\lambda$ =1 indicates their strict 2D behavior almost down to R=0. (b) Aslamazov-Larkin plot showing the behavior of the majority of the other samples, independent from their origin. We observe at higher temperatures a $\delta$ =1, that as the temperature approaches $T_c$, passes to a $\delta$ =0.5, indicating that the transition is 3D. (c) TEM image of crystals from the same batch as the sample described on Fig.~\ref{Res-3D}. Interferences are clearly seen at crystals surfaces, that can be interpreted as ordered stacking faults. Insert: Schema of stacking Bernal layers orderly intermingled with default ones \textit{x} (ABC-AB, Twisted graphene multilayers or other). (d) Upper panel: Electrical resistivity of MHS Aquadag grains showing a superconducting transition. Lower panel: Aslamazov-Larkin plot for the MHS Aquadag sample showing its 2D behavior.}
	\label{Aslamazov}
\end{figure}

\begin{figure}[ht]
	\centering
	\includegraphics[width=18cm]{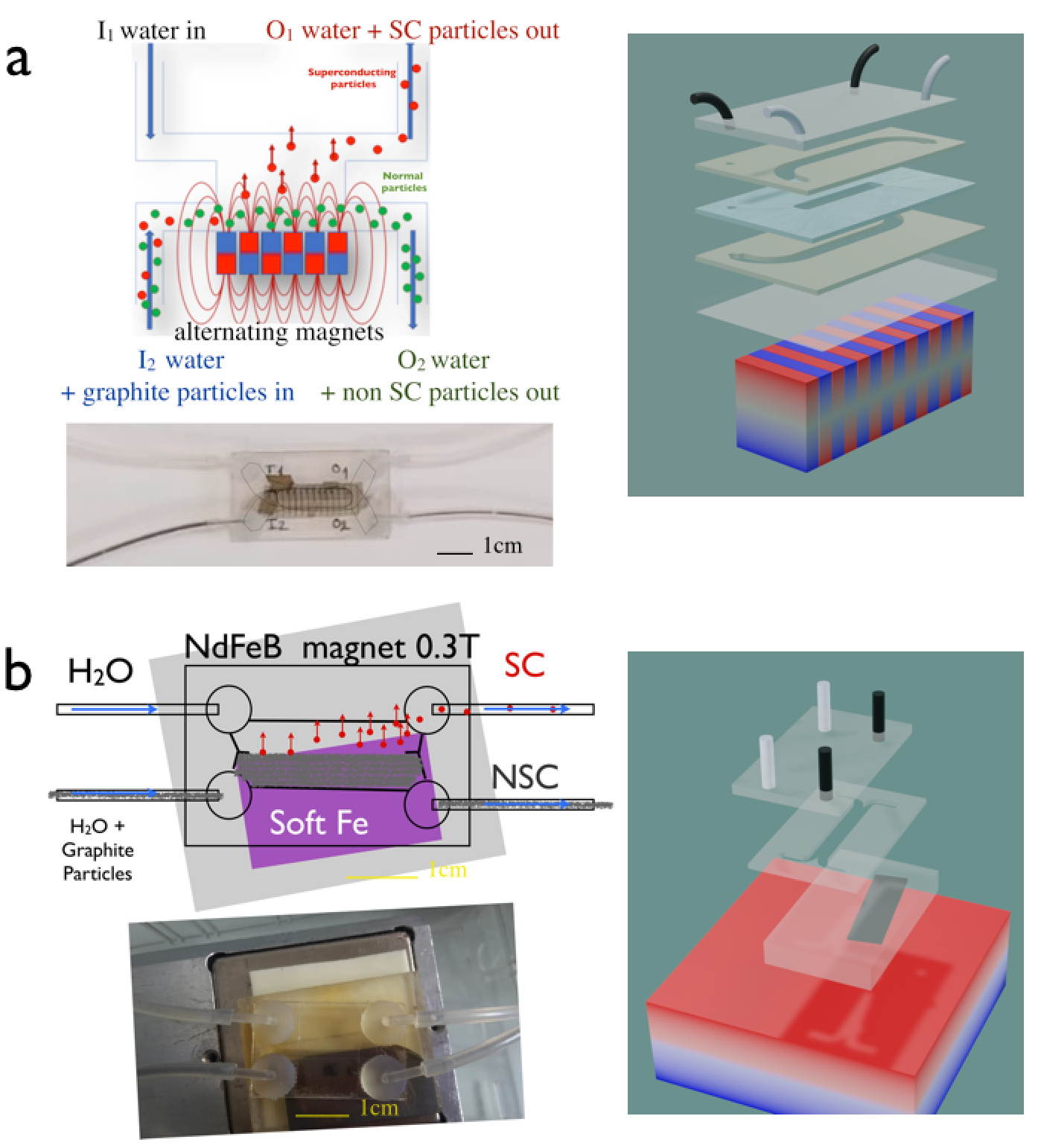}
	
	\caption {(a)  Magnetic vertically sorting (MVS) microfluidic device. Above left: Schema of the method with two circulating channels, one with suspended graphite particles below and the second on top only with water. On passing above the magnets the more diamagnetic superconducting particles change channel.  Below left: actual photo of the device. Right: 3D exploded view of the device.(b) Magnetic horizontally sorting microfluidic device (MHS). Above left: Schema of the device on top of a magnet with the two channels passing over a slanted soft iron foil. The superconducting particles are coin down by the magnetic gradient and forced to change channel. Below left: photo of the device. Right: 3D exploded view of the device.}
	\label{System_SI} 
\end{figure}

\begin{figure}[ht]
	\centering
	\includegraphics[width=11cm]{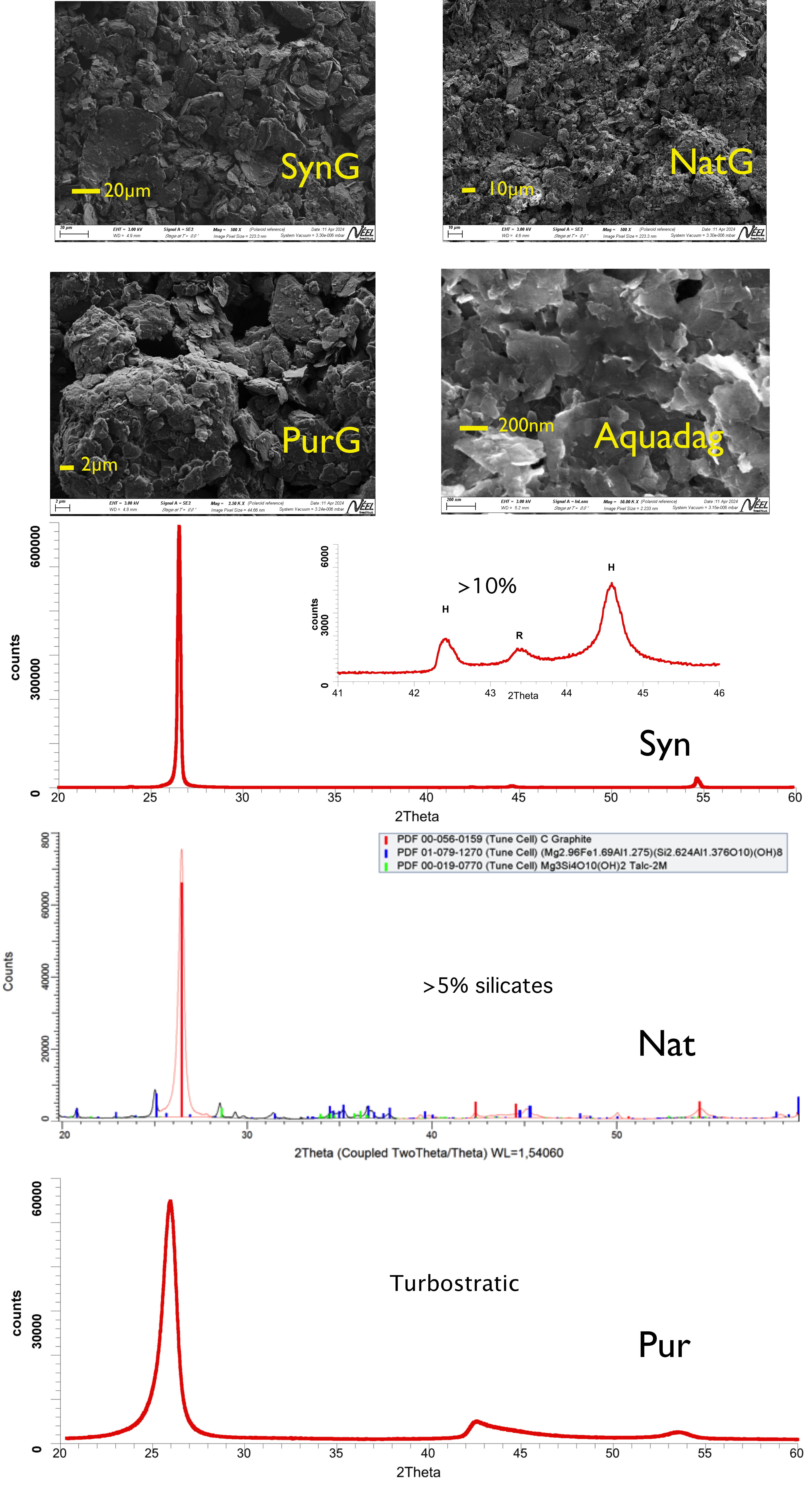}
	\caption{Upperpanel: Scaning electron images for SynG, NatG, PuG and Aquadag. Lower panel: X-rays spectra of the different graphite powders used in our study, before to magnetic sorting. First from top: Sigma-Aldrich  synthetic powder 282863 (SynG). The zoom from the X-rays spectra shows that a peak due to rhombohedral graphite is present and corresponds to $\ge$10\% of the sample. Second from top: Prolabo natural powder (NatG). Grain size is about $>$1$\mu$m. The X-rays spectrum shows that this powder has a rather high amont of different silicates due to its natural origin. Last from top panel: the purest available graphite, spectroscopically pure carbon rods from AGAR that were milled, with agathe balls and mortar, to obtain the needed powder(PurG). The grain size is about $>$1$\mu$m. The X-rays spectrum clearly shows that this graphite is turbostratic, i.e. carbon layers are arranged in a haphazardly folded or crumbled manner. Thus it has a lot of stacking defaults.}
	\label{X-rays} 
\end{figure}

\begin{figure}[ht]
	\centering
	\includegraphics[width=13cm]{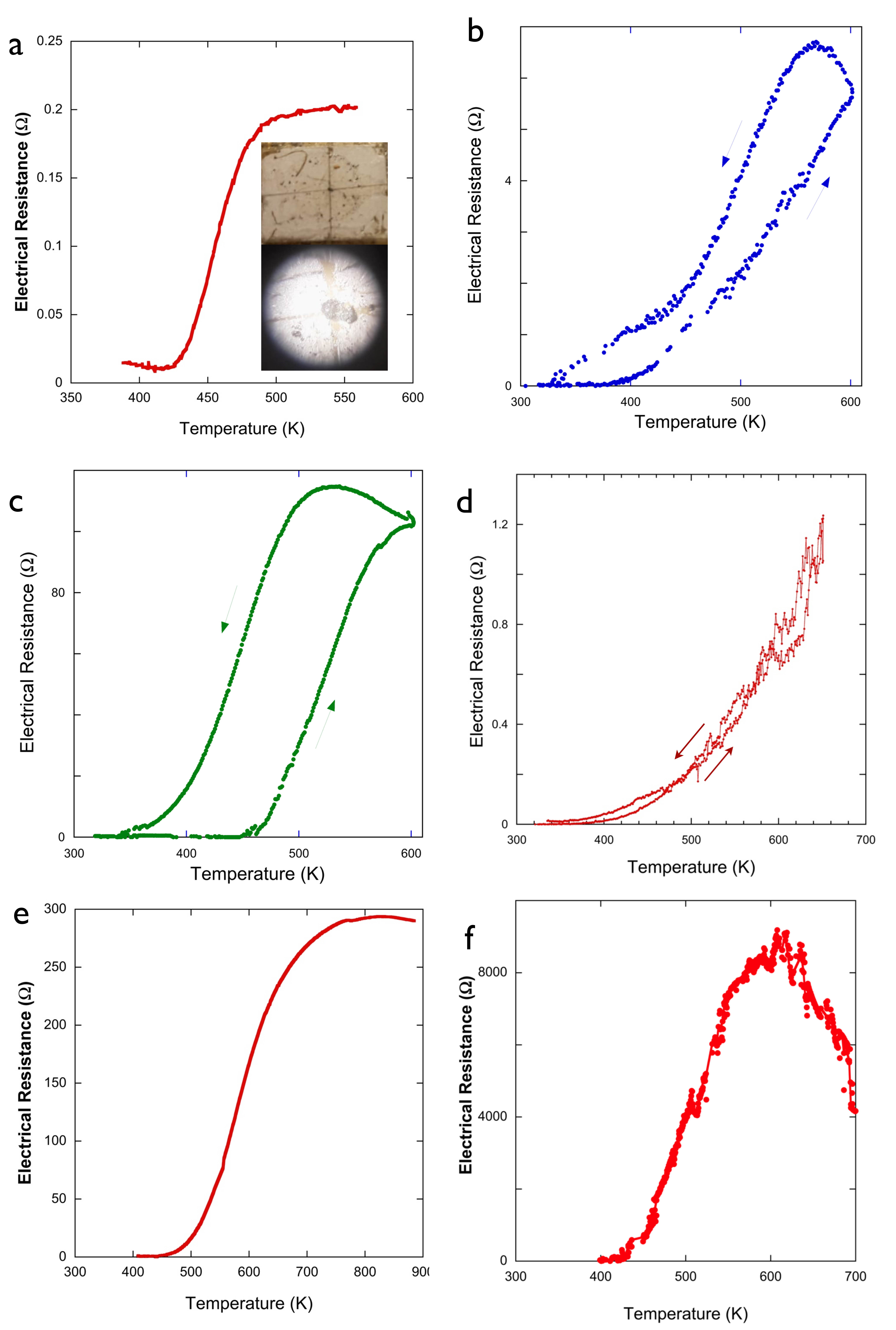}
	\caption{(a) Electrical resistance on a MDS SynG sample, showing a kink at about 600K and a full transition (albeit not to zero resistance) around 450K. This is the first observed transition in our study. Insert Upper panel: van der Pauw device used for this measurement made from spatulated silver epoxy with a channels incised with razor blade; the sample is the small dot in the middle. Lower panel: zoom sample made from agglomerated carbon particles on the crossing of the slits. Sample size $\sim$200$\mu$m. (b) Electrical resistance of a MVS NatG sample showing up and downwards transtionnhysteresis. (c) Electrical resistance of a MVS NatG sample showing up and downwards transtion hysteresis. (d)Electrical resistance of a MVS NatG sample showing up and downwards transtion hysteresis. (d) Electrical resistance of a MVS NatG sample showing up and downwards transtion hysteresis. At 650K we do not attain the normal state. 
		(e) Resistive transiton observed on a MHS PurG. The middle transition is just arond 580K. (f)  Resistive transiton observed on a MHS NatG}
	\label{Res-SI} 
\end{figure}

\begin{figure}[ht]
	\centering
	\includegraphics[width=14cm]{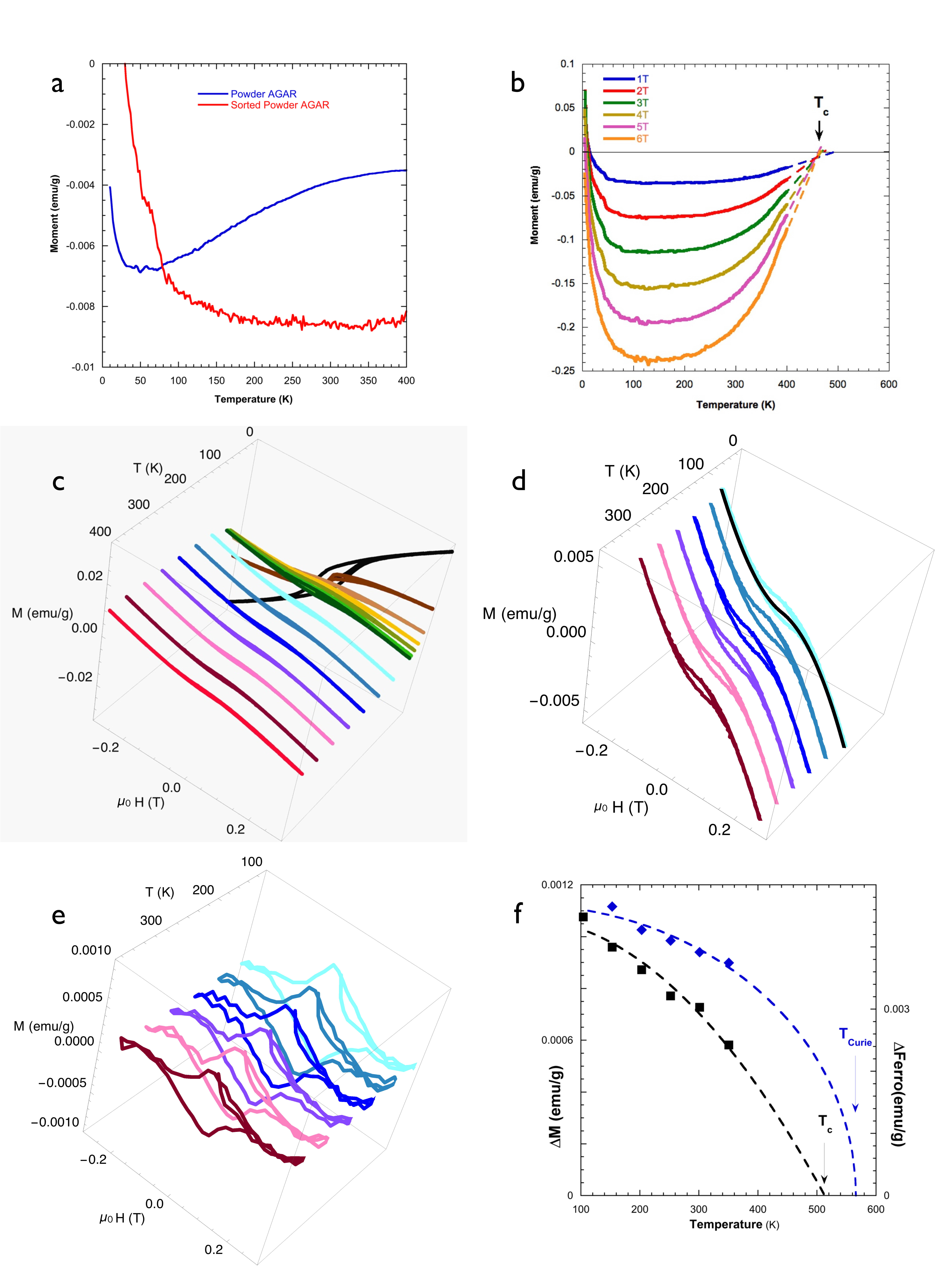}
	\caption{Magnetization results for MHS PurG sample . (a) Comparison of the ZFC magnetization of the virgin powder, obtained by milling of a AGAR graphite rod and the HS powder. We confirm the higher diamagnetism of the sorted powder at room temperature and a paramagnetic behavior at low temperatures, that in this sample should be intrinsic as it cannot be due to impurities. (b) ZFC magnetization measurements as a function of temperature at different fields. Its extrapolation at high temperatures allows a rough estimate of the T$_c\sim$460K. 3D plot of the magnetization cycles as a function of temperature, confirming the paramagnetic behavior at low temperatures. Detail of high temperature magnetization cycles. (e) Same cycles after substraction of a diamagnetic linear slope and a ferromagnetic cycle, showing what seem superconducting diamagnetic cycles. (f) Black squares : height of the precedent cycles as a function of temperature. Fitting the evolution with expression (1) yields a T$_c$=515$\pm$50K. We also plot the height of the ferromagnetic cycles (that should be intrinsic in this ultra-pure sample) as a function of temperature. Fitting with expression (2) we obtain a T$_{Curie}$=560$\pm$50K. The resistance of another sample of the same batch is shown in Fig.~\ref{Res-3D}(e), yielding the same T$_c$ within error bars.}
	\label{TDA1} 
\end{figure}

 We have also measured magnetization cycles as a function of magnetic field that we show on Fig.~\ref{TDA1}c. As with the precedent sample the low temperature cycles have a global paramagnetic dependence. While with increasing temperatures they pass to a  diamagnetic behavior, but with an unexpected s-shape ferromagnetic cycle superimposed on it (black line shown only on the 100K cycle on Fig.~\ref{TDA1}d), which can not be originated by iron impurities in this ultrapure sample. The cycle has an appreciable hysteresis. To analyze it, we fit a function composed of a sigmoidal ferromagnetism cycle and a diamagnetic linear dependence on it. We subtract the fitted function on each cycle and obtain the cycles shown on Fig.~\ref{TDA1}c. If we assume that the observed ferromagnetism is soft and its hysteresis negligeable, then we can suppose that they are essentially diamagnetic hysteretic cycles due to superconductivity. We extract the amplitude of the cycles at 0.07T, plot them as a function of temperature and fit them with expression (Eq. 1). We obtain a \textit{T$_c$}~ 550K$\pm$50K, compatible with the extrapolation on Fig.~\ref{TDA1}e. It is also is in good agreement with the \textit{T$_c$} obtained by the resistance measurement shown on Fig.~\ref{Res-3D}d for an MVS PurG sample. We can also study the fitted ferromagnetic cycles and plot their magnitude as a function of temperature and fit them with expression (Eq. 2). We obtain a \textit{T$_{Curie}$}=560$\pm$50K, that almost coincides with the value of the extracted superconducting \textit{T$_c$}. 
 
\textbf{Natural Graphite}:We continue with the magnetic measurements done on MHS NatG powder. The horizantal method is, by large, the more effective  of the three. Thus, the amount of obtained sample is then minimal, $<$ 0.1mg, but the superconducting percentage should be larger.
 
\begin{figure}[ht]
	\centering
	\includegraphics[width=15cm]{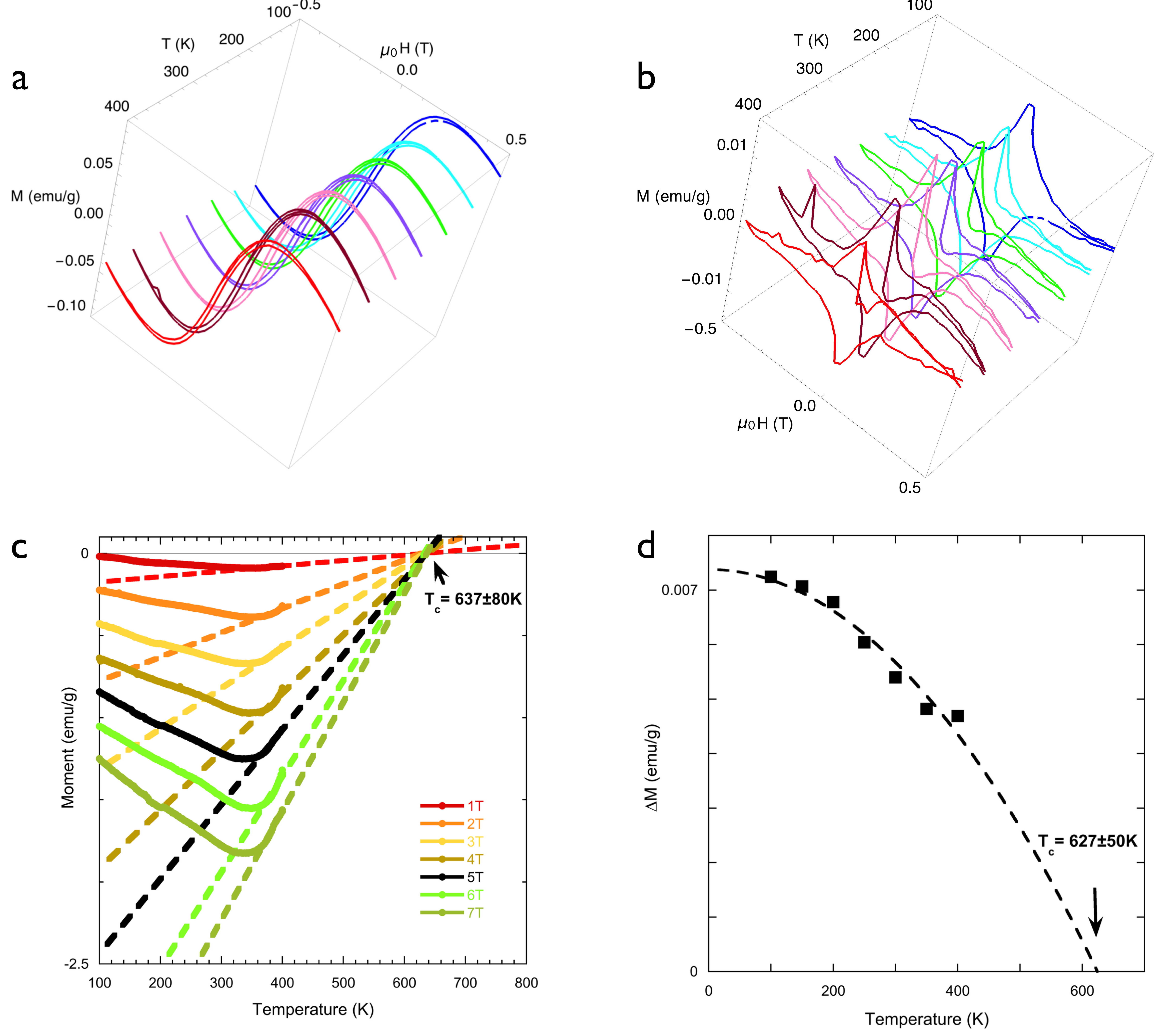}
	\caption{Magnetization results obtained on an extremely small ($\le$0.1mg) MHS NatG sample (a) Raw magnetization data showing the magnitude of the cycles as a function of temperature.The cycles have both a diamagnetic linear behavior and a ferromagnetic cycle superimposed, on what appears to be a diamagnetic hysteresis cycle. (b) Cycles obtained by subtracting  the fitted diamagnetic linear  and the ferromagnetic cycle behavior. We obtain  superconducting hysteretic cycles that diminish with increasing temperature. (c) Black squares : height of the precedent cycles as a function of temperature. Fitting the evolution with expression (2) yields a T$_c$=627$\pm$50K. (d) ZFC magnetization measurements as a function of temperature at different fields. Its extrapolation at high temperatures allows a rough estimate of the T$_c$=637$\pm$80K. The resistance of another sample of this batch is shown in Fig.~\ref{Res-3D}(a)and (b), yielding the same T$_c$ within error bars.
	}
	\label{Prolabo} 
\end{figure}

 The magnetization cycles obtained as a function of temperature are shown on Fig.~\ref{Prolabo}a. Although not shown, at low temperatures we obtain an antiferromagnetic Curie-like behavior, with \textit{T$_{Curie}$} $\sim$ -4K. In this case, though, the original graphite comes from a natural source and the non-sorted powder has many impurities. The cycles are extremely thin, but not empty. We can fit them with a diamagnetic linear and a sigmoidal dependence, which we substract, as done before. The obtained cycles are seen on Fig.~\ref{Prolabo}b, decreasing their amplitude with increasing temperature. Plotting the vertical span as a function of temperature and fitting the data with expression (Eq. 1), we obtain a \textit{T$_c$} =627$\pm$50K.
 
 Once more, due to the small amount of sample that we dispose, we perform a second determination of the transition temperature measuring several ZFC magnetization curves and determining where the curves converge at high temperature. We obtain a \textit{T$_c$} =637$\pm$80K, in good agreement with the previous value and with the T$_c$ determined from resistance measurements on another sample of the same batch Fig.~\ref{Res-3D}b.

\bibliographystyle{aipnum4-1}


\

\

\

 \

\

\end{document}